\documentclass[11pt,final,journal,letterpaper,oneside,reqno, onecolumn]{article}

\usepackage{graphicx} 
\usepackage{enumerate,graphics} 
\usepackage{color}
\usepackage{amsmath,amsfonts,amsthm,amssymb} 
\usepackage{ifthen}
\usepackage[caption=false,font=footnotesize]{subfig}
\usepackage{lineno}
\usepackage{fourier}
\usepackage{hyperref}
\usepackage{cite}
\usepackage{algpseudocode}
\usepackage{algorithm}
\usepackage{verbatim}
\usepackage[amssymb]{SIunits}
\usepackage{pstricks}

\usepackage{breakurl}
\usepackage{setspace}
\usepackage{multirow}

\usepackage[letterpaper,left=1.40in,right=1.40in,top=1.25in,bottom=1.25in]{geometry}

\newboolean{fig_pdf}
\setboolean{fig_pdf}{true}
\newboolean{fig_eng}
\setboolean{fig_eng}{true}

\DeclareMathOperator{\Var}{Var}

\begin{document}

\title{Interval edge estimation in SAR images}

\author{La\'{e}rcio~Dias,~Francisco~Cribari-Neto,~and~Raydonal~Ospina
\thanks{All authors are with the Departamento de Estat\'{\i}stica, Universidade Federal de Pernambuco, Cidade Universit\'{a}ria, Recife/PE, 50740--540, Brazil.}
 }
        
\maketitle

\begin{abstract}
This paper considers edge interval estimation between two regions of a Synthetic Aperture Radar (SAR) image which differ in texture. This is a difficult task because SAR images are contaminated with speckle noise. Different point estimation strategies under multiplicative noise are discussed in the literature. It is important to assess the quality of such point estimates and to also perform inference under a given confidence level. This can be achieved through interval parameter estimation. To that end, we propose bootstrap-based edge confidence interval. The relative merits of the different inference strategies are compared using Monte Carlo simulation. The results show that interval edge estimation can be used to assess the accuracy of an edge point estimate. They also show that interval estimates can be quite accurate and that they can indicate the absence of an edge. In order to illustrate interval edge estimation, we also analyze a real dataset. 

\end{abstract}



\section{Introduction}\label{S:introduction}

Imagery obtained through the use of coherent illumination suffers from a noise known as speckle. This is the case, for instance, of synthetic aperture radar (SAR) images \cite{oliverquegan2004}. The noise is usually modeled in a multiplicative fashion and different inference strategies are used in order to extract useful information from data subject to such speckle noise. Such sensors are particularly useful because they do not require external sources of illumination and their wavelength is not affected by weather conditions. 

There are different types of texture for a given region of a SAR image. Textures range from homogeneous (i.e., surfaces with little texture such as lakes, deforestation areas, crop fields, deserts and areas covered by snow), to extremely heterogeneous (i.e., areas that have very intense texture, such as urban areas) with heterogeneous targets, as forests, in between. Such a classification depends on several factors: signal frequency, polarization, and angle of incidence, among others (see \cite{oliverquegan2004}, section 13.2.3).

One of the most important image processing tools is edge detection, which allows one to detect the position of the edge between two regions. Many edge detection strategies that have been proposed in the literature yield edge point estimates. Our main goal in this paper is to consider interval edge estimation. We introduce different strategies that can be used to produce a confidence interval for an edge that separates two regions of a SAR image with different textures. The interval estimators are obtained by setting the desired confidence level and their lengths are indicative of how certain one can be that the edge has been located. Interval edge estimation yields important information since point edge estimation can be a difficult task due to speckle noise and since it can signal the absence of an edge, as we shall see.

Approaches of several kinds for detecting an edge between neighboring regions with different textures have been proposed in the literature. Nascimento et al.\ \cite{nascimento2013} used stochastic entropies and distances as a criterion for locating edges. Baselice and Ferraioli \cite{2012.03.Baselice.Ferraioli} carried out edge detection based on the ground properties of urban areas using Markov chains to jointly model the amplitudes and interferometric phases of two SAR complex images. Fu et al.\ \cite{2012.11.Fu} proposed an edge detector which uses `square successive differences of averages' as an indicator of `edge strength'. Singhal and Singh \cite{2011.11.Singhal.Singh} locate edges by removing of speckle noise using mathematical morphology. Alonso et al.\ \cite{2011.01.Alonso} use a robust and unsupervised edge enhancement algorithm based on a combination of wavelet coefficients at different scales. Frery et al.\ \cite{frery2010} consider polarimetric SAR image region boundary detection based on B-spline active contours. Oliver et al.\ \cite{oliver1996} propose a maximum likelihood method that can be used to detect an edge within a window and determine its location. Touzi et al.\ \cite{touzi} provide a constant-false-alarm-rate (CFAR) edge detector based on the ratio between pixel values. Fj{\o}rtoft et al.\ \cite{fjortoft} propose a step-edge
detector which is optimal in the minimum mean
square error under a stochastic multiedge model. Melgar et al. \cite{melgar} evaluate and compare different edge-detection algorithms developed for high-speckle SAR images. Baselice et al.\ \cite{baselice} provide an approach based on the exploitation of real and imaginary parts of single-look complex acquired data. Their technique is developed exploiting Markov random fields. Qian at al.  \cite{Qian} propose an  automatic local thresholding algorithm (ALTA)  to determines threshold and improve ratio-typical synthetic aperture radar (SAR) edge detector.

Gambini et al.\ \cite{Gambini2006, Gambini2008} proposed a parametric point estimator (i.e., detector) for the edge location using an objective function that is maximal at the transition points that lie on strip of pixels. Nonparametric detectors have also been proposed. Bovik et al.\ \cite{Bovik1986} used nonparametric methods to detect edges under additive Gaussian noise. Beauchemin et al.\ \cite{Beauchemin1998} used an alternative approach based on the  Wilcoxon-Mann-Whitney statistic to detect changes in adjacent sets of pixels. Lim and Jang \cite{Lim2002} used two sample tests to detect edges in images subject to noise. Girón et al.\ \cite{Giron2012}, following the main ideas in \cite{Gambini2006}, introduced detectors based on the Mann-Whitney, squared ranks to variances, Kruskal-Wallis and empirical distribution nonparametric test statistics. The authors used Monte Carlo simulation to compare the small sample performances of their detectors to that of Gambini's estimator. Their numerical evidence shows that the Kruskal-Wallis and Gambini's detectors behave similarly, the former however being considerably less costly from a computational viewpoint. Indeed, their numerical results show that edge detection using the Kruskal-Wallis detector is approximately one thousand times faster than that based on Gambini's maximum likelihood detector. 

We chose to construct confidence intervals based on the Kruskal-Wallis point estimator because the available numerical evidence shows that it performs as well as the maximum likelihood detector (Gambini's detector) at a much lower computational cost. We use bootstrap resampling in order to obtain an estimate of its distribution, which is unknown. The bootstrap methods usually delivers accurate estimates \cite{DavisonHinkley}. We also consider different bootstrap interval estimators: basic bootstrap, percentile and two variants percentile-$t$ method that we propose. For details on bootstrap interval estimation, the reader is referred to Efron and Tibishirani \cite{Efron1993}.

The remainder of the article unfolds as follows. Section~\ref{S:model} introduces the statistical model we shall use to model SAR data. Section~\ref{S:edgedetection} presents the Gambini \cite{Gambini2006,Gambini2008} and Kruskal-Wallis detectors. Section~\ref{S:nonparametricbootstrap} considers edge bootstrap interval estimation. Section~\ref{S:montecarlo} presents Monte Carlo evidence. Section~\ref{S:application} contains an application that uses real (not simulated) data. Finally, Section~\ref{S:conclusion} offers some concluding remarks.


\section{The SAR image model}\label{S:model}

The multiplicative statistical model is widely used in SAR image data analyses. According to this model, the data 
are described by a random variable $Z$ which can be viewed as the product of the independent random variables $X$ and $Y$, where $X$ models the properties of the imaged area (backscatter) and 
$Y$ models the multiplicative noise (speckle) introduced by the use of coherent illumination which degrades the image quality. 

According to the model proposed by Frery et al.\ \cite{Frery1997}, the speckle noise is gamma distributed, denoted by $Y \sim \Gamma(L,L)$, and the backscatter can be modeled using the inverse gamma distribution, $X \sim \Gamma^{-1}(\alpha, \gamma)$. Thus, $Z=XY$ is distributed as ${\mathcal{G}}^0_{\mathcal{I}}$, $Z \sim {\mathcal{G}}^0_{\cal{I}}(\alpha, \gamma, L)$, whose density function is 
\begin{equation}
f_{Z}(z) = \frac{L^L \Gamma(L-\alpha)}{\gamma^\alpha \Gamma(L) \Gamma(-\alpha)} \frac{z^{L-1}}{(\gamma+Lz)^{L-\alpha}} , \mbox{\hspace{0.5cm}} L \geq 1, -\alpha, \gamma, z > 0.
\label{eq:densidade.Z}
\end{equation}

Images represented in the intensity format can be described by the distribution ${\mathcal{G}}^0_{\mathcal{I}}$, denoted by $Z \sim {\mathcal{G}}^0_{\mathcal{I}}$. The parameters that index such a distribution are: (i) the number of looks $L\geq 1$, which is a measure of the signal-to-noise ratio, (ii) the scale parameter $\gamma>0$, which is related to the relative strength between the incident and reflected signals, and (iii) the roughness parameter $\alpha<0$, which relates to the land type texture. The larger the value of $\alpha$, the more heterogeneous the area: when $\alpha < -10$ the area is very homogeneous (e.g., pastures), when $-10 < \alpha < -4$ the area is heterogeneous (e.g., forests) and when $-4 < \alpha < 0$ the area is extremely heterogeneous (e.g., urban areas). In what follows we shall assume that the number of looks is known.

Using (\ref{eq:densidade.Z}), it can be easily shown that the $r$th noncentral moment is 
\begin{equation}
E[Z^r] = \left(\frac{\gamma}{L}\right)^r \frac{\Gamma(-\alpha-r)\Gamma(L+r)}{\Gamma(-\alpha) \Gamma(L)},
\label{eq:momento.r}
\end{equation}
if $-\alpha > r$, and $\infty$ otherwise. Notice that the value of $\gamma$ that corresponds to $E[Z]=1$ is 
\begin{equation}
\gamma = \frac{\Gamma(-\alpha) \Gamma(L) L}{\Gamma(-\alpha-1)\Gamma(L+1)}.
\label{eq:gamma.norm}
\end{equation}

Gambini's edge detection method requires estimation of $\alpha$ and $\gamma$. Several estimators for these parameters are available in the literature. For instance, \cite{Bustos2002} and \cite{Allende2006} consider robust paramater estimation, and \cite{Silva2008,klaus2005,Cribari2002} propose bias-corrected estimators. The work \cite{Frery2004} numerically evaluate the accuracy of parameter estimation in small samples. We shall use a Kruskal-Wallis nonparametric detector which requires no distribution assumption and also no parameter estimation.

\section{Edge detection in SAR imagery}\label{S:edgedetection}

\subsection{The Gambini parametric detector.} The Gambini algorithm is based on the fact that if a point belongs to the edge, then a sample taken from its neighborhood is expected to exhibit a noticeable change in the parameter values which are used to describe the pixel distribution on either side of the edge. Let $s$ be a line segment (detection line) such that, in principle, one point that belongs to $s$ also belongs to the edge that separates two regions. 
The interest lies in determining which point detection line is a transition boundary. The segment $s = (z_1, \ldots, z_N)$ is a strip of pixels obtained by discretization of a straight line on the image.
Assume that there is a nonnull intersection between the edge and the detection line and let $z_j$ be a point such that $(z_1, \ldots, z_j)$ comes from a ${\cal{G}}^0_{\cal{I}}$ distribution with parameters $(\alpha_\ell, \gamma_\ell)$ and $(z_{j+1}, \ldots, z_N)$ comes from another ${\cal{G}}^0_{\cal{I}}$ distribution with parameters $(\alpha_r, \gamma_r)$. Then, the value of $j$ indicates the edge location. Be $L$ the number of looks. To find the transition point at $s$, consider the likelihood function given by
\begin{align}
&1\!\!{\rm L} = 1\!\!{\rm L}(\alpha_\ell, \gamma_\ell, \alpha_r, \gamma_r)= \prod_{k=1}^{j} Pr(z_k;\alpha_\ell, \gamma_\ell)\\ &\times \prod_{k=j+1}^{N} Pr(z_k;\alpha_r, \gamma_r). 
\label{eq:verossimilhanca.amostral}
\end{align}
The log-likelihood function 
\begin{equation*}
\mathcal{L} = \ln 1\!\!{\rm L} = \sum_{k=1}^{j} \ln f_{ {\cal{G}}^0_{\cal{I}} }(z_k;\alpha_\ell, \gamma_\ell) + \sum_{k=j+1}^{N} \ln f_{ {\cal{G}}^0_{\cal{I}} }(z_k;\alpha_r, \gamma_r),
\end{equation*}
where $j \in \left\{ 1, \ldots, N-1 \right\}$, is maximized when the index $j$ is the edge. 
Using Equation (\ref{eq:densidade.Z}), it follows that 
\begin{eqnarray}
\mathcal{L} = \sum_{k=1}^{j} \ln \frac{L^L \Gamma(L-\hat{\alpha}_\ell) z^{L-1}_k} { \hat{\gamma}_\ell^{\hat{\alpha}_\ell} \Gamma(L) \Gamma(-\hat{\alpha}_\ell)(\hat{\gamma}_\ell + Lz_k)^{L-\hat{\alpha}_\ell}}\nonumber \\
+ \sum_{k=j+1}^{N} \ln \frac{L^L \Gamma(L-\hat{\alpha}_r) z^{L-1}_k} { \hat{\gamma}_r^{\hat{\alpha}_r} \Gamma(L) \Gamma(-\hat{\alpha}_r)(\hat{\gamma}_r + Lz_k)^{L-\hat{\alpha}_r}}.
\label{eq:log-verossimilhanca}
\end{eqnarray}
Thus, the estimated transition point at $s$ is given by
\begin{equation}
\hat{\jmath}_{GE} = \arg\max_j \mathcal{L}.
\label{eq:borda.est}
\end{equation}

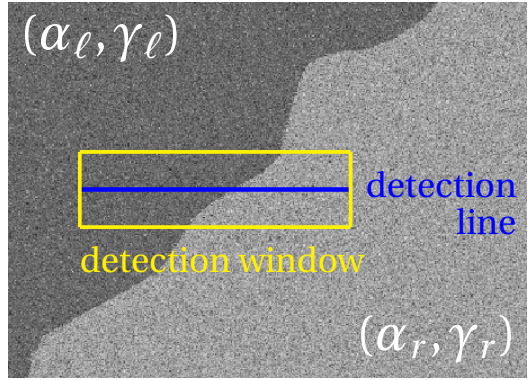
\begin{figure}[!h]
  \begin{center}
	\begin{pspicture}(0,0)(5,5)
		\rput(2.5,2.5){\includegraphics[height=5cm]{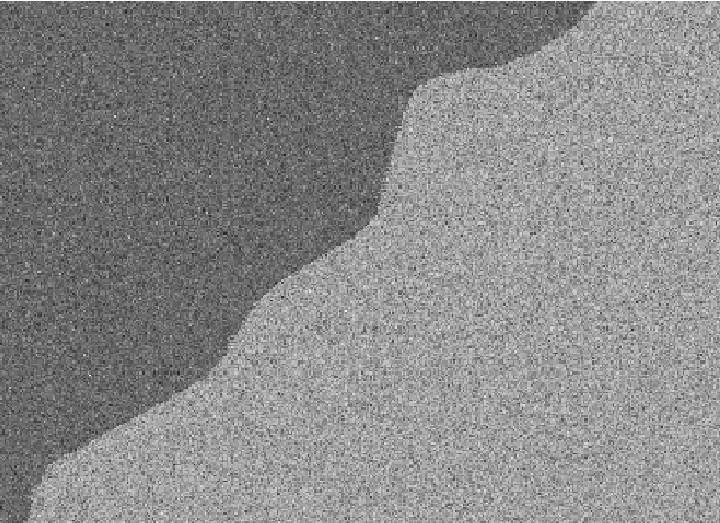}}
		\rput[bl](-0.8,4.2){\huge {\white $(\alpha_\ell,\gamma_\ell)$}}
		\rput[br](5.8,0.2){\huge {\white $(\alpha_r,\gamma_r)$}}
		\rput[br](5.8,2.4){\Large \textcolor{blue}{detection}}
		\rput[br](5.8,1.9){\Large \textcolor{blue}{line}}
		\psline[linewidth=0.06,linecolor=blue](0,2.5)(3.6,2.5)
		\psline[linewidth=0.05,linecolor=yellow](0,2)(3.6,2)
		\psline[linewidth=0.05,linecolor=yellow](0,3)(3.6,3)
		\psline[linewidth=0.05,linecolor=yellow](0,2)(0,3)
		\psline[linewidth=0.05,linecolor=yellow](3.6,2)(3.6,3)
		\rput[bl](0,1.4){\Large \textcolor{yellow}{detection window}}
	\end{pspicture}
    \caption{Using pixels in a detection window (yellow), one can, using the Gambini estimator, locate the edge between two distinct regions along a detection line (blue).}
    \label{fig.motivacao}
  \end{center}
\end{figure}

Edge detection using the Gambini estimator uses information on the pixels that lie in a neighborhood of a rectangular region (detection window) around detection line. Figure~\ref{fig.motivacao} shows a simulated SAR image that contains two distinct regions and also a detection window and an inner detection line.

The Gambini's estimator $\hat{\jmath}_{GE}$ requires the estimation of $(\alpha_\ell, \gamma_\ell)$ and $(\alpha_r, \gamma_r)$ for all possible values of $j \in \left\{ 1, \ldots, N-1 \right\}$. Such estimations are carried out using rectangular windows in which the detection line coincides with the major axis of the window. Note that for extreme values of $j$ estimation is performed using a small sample, and as a consequence the resulting estimate can be poor. The papers \cite{Silva2008} and \cite{Cribari2002} present estimation approaches that can yield more accurate inferences.

\subsection{Nonparametric edge detection.} Edge detection using nonparametric methods as alternative to the Gambini's estimator was proposed by Girón et al.\ \cite{Giron2012}. In what follows we shall use their best performing estimator, namely: Kruskal-Wallis. 

Consider $k$ independent random samples (with possibly different sizes) from $k$ identical populations or from $k$ populations such that at least one of them tends to produce observations with larger values, the $i$th sample containing $n$ observations. Let $N=\sum^k_{i=1}n_i$ be the total number of observations. The Kruskal-Wallis statistic was developed to test the null hypothesis that  the $k$-samples come from the same population. If the null hypothesis is false, then at least one of the $k$ samples will tend to produce observations with larger values. In order to compute the Kruskal-Wallis statistic we have to consider $N$ observations and assign ranks to them. Let $R(z_{ij})$ be the rank of $z_{ij}$, the $j$th observation of the $i$th sample, and let $R_i = \sum^{n_i}_{j=1}R(z_{ij})$ be the sum of ranks to the $i$th sample. The test statistic is given by
$$
T = \frac{1}{S^2}\left(\sum^k_{i=1}\frac{R_i^2}{n_i}-\frac{N(N+1)^2}{4}\right),
$$
where
$$
S^2 = \frac{1}{N-1}\left(\sum_{i=1}^k\sum_{j=1}^{n_i} R(z_{ij}) - \frac{N(N+1)^2}{4}\right).
$$

If the number of ties is sufficiently small (or null), it can be shown that the statistic $T$ can be expressed as 
\begin{equation}
T_{KW} = \frac{12}{N(N+1)}\sum^k_{i=1}\frac{R_i^2}{n_i}-3(N+1).
\label{eq:funcaoObjetivo.kw}
\end{equation}

Edge estimation using $T_{KW}$ is straightforward. Consider a strip of pixels $s = (z_1, \ldots, z_N)$ and assume that the $j$th element is the edge. The edge thus splits the data into two samples, $(z_1, \ldots, z_j)$ and $(z_{j+1}, \ldots, z_N)$, whose sample sizes are  $n_1 = j$ and $n_2 = N-j$, respectively, where $j \in \{1,\ldots,N-1\}$. We conclude that the two samples come from different populations when $T_{KW}$ is large, the corresponding edge estimator being
\begin{equation}
\hat{\jmath}_{KW} = \arg\max_j T_{KW}.
\label{eq:est.kw}
\end{equation}
This edge estimator is computationally less costly than Gambini's parametric estimator.

\section{Bootstrap-based edge confidence intervals}\label{S:nonparametricbootstrap}

It is important for practitioners to be able to compute not only edge point estimates but also edge interval estimates, that is, confidence intervals for SAR image edges. We note that the distribution of KW point estimator is unknown and hence the usual corresponding confidence intervals cannot be obtained. An alternative is to use bootstrap resampling to construct interval estimates \cite{DavisonHinkley, Efron1993, bootstrap3}. The main idea is to use data resampling to construct pseudo-samples, to compute the point estimate using each artificial sample and then to use all point estimates to produce a confidence interval.

In what follows we shall consider bootstrap edge interval estimation. In particular, the following bootstrap interval estimators shall be considered: boot\-strap basic method (BBM), percentile (PERC), and Studentized (or percentile-$t$, boot\-strap-$t$) (ST). Given the high computational burden of the percentile-$t$ method, we propose two new variants of such a confidence interval that are much less computer-intensive.

Consider again a strip of pixels $s = (z_1, \ldots, z_N)$, with an edge between two regions located at $j \in \{1,\ldots,N\}$. Let $\hat{\jmath}_{kw}$ be the KW edge estimate. Using the estimate $\hat{\jmath}_{kw}$ we can obtain pseudo-samples $s^{*} = (z^{*}_1, \ldots, z^{*}_N)$, where $(z^{*}_1, \ldots, z^{*}_{\hat{\jmath}_{kw}})$ is obtained by randomly sampling with replacement from $(z_1, \ldots, z_{\hat{\jmath}_{kw}})$. Similarly, $(z^{*}_{\hat{\jmath}_{kw}+1}, \ldots, z^{*}_N)$ is obtained by randomy sampling with replacement from $(z_{\hat{\jmath}_{kw}+1}, \ldots, z_N).$ Let $\hat{\jmath}^{*}$ be the edge estimate obtained by resampling a strip of pixels $s^{*}$. After executing this scheme $B$ times, we obtain $B$ edge estimates ($\hat{\jmath}^{*}_1,\ldots,\hat{\jmath}^{*}_B$). This subsampling process is known as nonparametric bootstrap. Let $\hat{\jmath}^{*}_{(1)},\ldots,\hat{\jmath}^{*}_{(B)}$ the ordered values of  $\hat{\jmath}^{*}_1,\ldots,\hat{\jmath}^{*}_B$. The empirical distribution of $\hat{\jmath}_{KW}$ is given by
\begin{equation}
\widehat{F}_B^{*}(j') = \frac{ \# \{ \hat{\jmath}^{*}_{(b)} \leq j' \} } {B},\quad b \in \{1,\ldots,B\}.  
\label{eq:emp.dist.j.kw}
\end{equation}
It can be used for constructing confidence intervals for the true edge location.

Assuming that $\hat{\jmath}_{KW}$ is a consistent estimator of the edge and given that the true distribution of $\hat{\jmath}_{KW}$ is unknown, one can construct an edge confidence interval based on the bootstrap approximation to the distribution of $\hat{\jmath}_{KW} - j.$ Here, the quantiles are obtained by using the ordered values of $\hat{\jmath}^{*}_{KW} - \hat{\jmath}_{kw}$. The $100(1-a)\%$ basic bootstrap method (BBM) confidence interval \cite{DavisonHinkley} is 
\begin{align}
&\left[\hat{\jmath}^{*}_{a/2},\hat{\jmath}^{*}_{1-a/2}\right]_{BBM} = \nonumber \\ &\left[\hat{\jmath}_{kw} - (\hat{\jmath}^{*}_{(B(1-a/2))} - \hat{\jmath}_{kw}), \hat{\jmath}_{kw} - (\hat{\jmath}^{*}_{(B(a/2))} - \hat{\jmath}_{kw})\right].
\label{eq:ic.bbm}
\end{align}

An alternative bootstrap confidence interval is the Percentile interval. The $100(1-a)\%$ percentile interval \cite{DavisonHinkley} is given by 
\begin{equation}
\left[\hat{\jmath}^{*}_{a/2},\hat{\jmath}^{*}_{1-a/2}]_{PERC} = [\hat{\jmath}^{*}_{(B(a/2))}, \hat{\jmath}^{*}_{(B(1-a/2))}\right].
\label{eq:ic.perc}
\end{equation}
It is noteworthy that this interval may be asymmetric and only contains proper values of the edge. 

An alternative approach is to construct a bootstrap studentized confidence interval: the percentile-$t$ or bootstrap-$t$ interval. To that end, in each bootstrap replication we compute the quantity
\begin{equation}
Z^{*}_{b} = \frac{\hat{\jmath}^{*}_{b} - \hat{\jmath}_{kw}}{\hat{v}^{*1/2}_{b}},
\label{eq:studentizacao}
\end{equation}
where $\hat{v}^{*1/2}_{b}$ is the standard error $\hat{\jmath}^{*}$ obtained in the $b$th bootstrap replication. The $a$th sample quantile, $\hat{t}_{(a)}$, is obtained as follows: 
$$
\frac{\#\{ Z^{*}_{b} \leq \hat{t}_{(a)}\}}{B} = a,
$$
$b\in\{1,\ldots\,B\}$. The $100(1-a)\%$ studentized bootstrap confidence interval \cite{DavisonHinkley} is then given by
\begin{equation}
\left[\hat{\jmath}^{*}_{a/2},\hat{\jmath}^{*}_{1-a/2}]_{ST} = [\hat{\jmath}_{kw} - \hat{v}^{1/2}\hat{t}_{(B(1-a/2))},\hat{\jmath}_{kw} - \hat{v}^{1/2}\hat{t}_{(B(a/2))}\right].
\label{eq:ic.st}
\end{equation}

The sample variance $v^{*}_{b}$ can be computed by using a sub-bootstrap (i.e., a second level bootstrap) with $B'$ replications, where $B' = 50$ is the minimum number of resamples that yields reasonable estimates. Efron and Tibshirani \cite{Efron1993} recommend using $B' = 200$. However, they also recommend using $B=1000$. This implies that estimation of $v^{*}_{b}$ using a second level bootstrap adds considerable computational burden, even when one uses $B' = 50.$

In what follows we shall propose two alternative approximate bootstrap-$t$ interval estimators, denoted ST1 and ST2. By using them one can compute $v^{*}_{b}$ without having to resort to a second level bootstrap. 

The idea behind out ST1 bootstrap method is to estimate $v^{*}_{b}$ using a subset of size $B'$ from the replications $\hat{\jmath}^{*}_1,\ldots,\hat{\jmath}^{*}_B$ imposing the restriction that the sample variance is positive. Let $\hat{\jmath}^{*'} = \{\hat{\jmath}^{*'}_{1},\ldots,\hat{\jmath}^{*'}_{B'}\}$ be the randomly selected subset of $B$ bootstrap estimates. Our algorithm can be outlined as follows: 

\begin{algorithm}[!h]
\caption{ST1 Method}\label{alg:st1}
\begin{algorithmic}[1]
\If{$\hat{\jmath}^{*}_{b} = \hat{\jmath}_{kw}$}
  \State $Z^{*}_{b} \leftarrow 0$
\Else
\If{$\Var{(\hat{\jmath}^{*}_1,\ldots,\hat{\jmath}^{*}_B)}\neq 0$}
  \State Obtain $\hat{\jmath}^{*'} = \{\hat{\jmath}^{*'}_{1},\ldots,\hat{\jmath}^{*'}_{B'}\}$ by drawing with replacement from  $\{\hat{\jmath}^{*}_1,\ldots,\hat{\jmath}^{*}_B\}$.
  \State $\hat{v}^{*}_{b} \leftarrow \Var{(\hat{\jmath}^{*'})}$
  \If{$\hat{v}^{*}_{b} = 0$}
    \State Go back to line $5$ (do this at most $B''>0$ times).
  \EndIf
  \If{line $8$ was repeated $B''$ times}
    \State $\hat{v}^{*}_{b} \leftarrow \Var{(\hat{\jmath}^{*}_1,\ldots,\hat{\jmath}^{*}_B)}$
  \EndIf
\Else
  \Repeat 
  \State Obtain $\hat{\jmath}^{*'}_{*}$ by estimating the edge using a new bootstrap replication of pixels strip $s$.
  \Until{$\hat{\jmath}^{*'}_{*} \neq \hat{\jmath}^{*}_1$ (do this at most $B''>0$ times)}
  \If{$\hat{\jmath}^{*'}_{*} = \hat{\jmath}^{*}_1$}
    \State $\hat{\jmath}^{*'}_{*} \leftarrow \hat{\jmath}^{*}_1 + 1$
    \State $\hat{v}^{*}_{b} \leftarrow \Var{(\hat{\jmath}^{*}_{1},\ldots,\hat{\jmath}^{*}_{B'-1}, \hat{\jmath}^{*'}_{*})}$
  \EndIf
\EndIf
\EndIf
\end{algorithmic}
\end{algorithm}

It is noteworthy that Algorithm \ref{alg:st1} must be executed for each $b \in \{1,\ldots,B\}$. This algorithm is considerably less computationally intensive than the standard bootstrap-$t$ algorithm in which a second level bootstrapping scheme is used for variance estimation. 

At Algorithm \ref{alg:st2} is introduced a second approximation to the bootstrap-$t$ method.

\begin{algorithm}[!h]
\caption{ST2 Method}\label{alg:st2}
\begin{algorithmic}[1]
\State Obtain estimates $\hat{\jmath}^{*}_{x} \leftarrow \{\hat{\jmath}^{*}_{B+1},\ldots,\hat{\jmath}^{*}_{B+B_{x}}\}$ of the edge based on $B_{x}$ new resamples from the pixels strip $s$.
\If{$\Var{(\hat{\jmath}^{*}_{x})} \neq 0$}
  \State Obtain $\hat{\jmath}^{*'} = \{\hat{\jmath}^{*'}_{1},\ldots,\hat{\jmath}^{*'}_{B'}\}$ ($B'<B_{x}$) by drawing with replacement from $\hat{\jmath}^{*}_{x}$.
  \State $\hat{v}^{*}_{b} \leftarrow \Var{(\hat{\jmath}^{*'})}$.
  \If{$\hat{v}^{*}_{b} = 0$}
    \State Go back to line $3$. This step is to be executed at most $B''$ times $(0<B''<B_{x})$.
  \EndIf
  \If{line $6$ was repeated $B''$ times}
    \State $\hat{v}^{*}_{b} \leftarrow \Var{(\hat{\jmath}^{*}_{x})}$.
  \EndIf
\Else
  \Repeat
    \State Obtain $\hat{\jmath}^{*'}_{*}$ by estimating the edge using a new bootstrap replication of pixels strips $s$.
  \Until{$\hat{\jmath}^{*'}_{*} \neq \hat{\jmath}^{*}_{B+1}$ (do this at most $B''>0$ times)}
  \If{$\hat{\jmath}^{*'}_{*} = \hat{\jmath}^{*}_{B+1}$}
    \State $\hat{\jmath}^{*'}_{*} \leftarrow \hat{\jmath}^{*}_{B+1} + 1$
  \EndIf
  \State $\hat{v}^{*}_{b} \leftarrow \Var{(\{\hat{\jmath}^{*}_{B+1},\ldots,\hat{\jmath}^{*}_{B+B'-1}, \hat{\jmath}^{*'}_{*}\})}$
\EndIf
\end{algorithmic}
\end{algorithm}

Notice that ST1 is less computationally expensive that the ST2 and that they are both less computationally costly than the standard bootstrap-$t$, in which a second level (inner) bootstrap is carried out for variance estimation.

\section{Numerical results and discussion}\label{S:montecarlo}

In what follows we shall report the results of several Monte Carlo simulations which were performed to assess the finite sample merits of different edge interval estimates in SAR images. All simulations were carried out using \textsc{Ox} matrix programming language \cite{Doornik2002}. The hardware used was an Intel(R) Core(TM)2 Quad CPU Q6600 2.40\giga\hertz\ computer running on Ubuntu Linux. Graphics were produced using \textsc{R} programming environment \cite{Venables2002}.

The simulated data have been generated according to the statistical distribution ${\mathcal{G}}^0_{\mathcal{I}}$ given in \eqref{eq:densidade.Z} and random number generation can be easily performed. It suffices to use the fact if $W \sim \Gamma^{-1}(k,\theta)$ then $1/W  \sim \Gamma(k,1/\theta)$. Thus, in order to generate a pseudo-random number from  ${\cal{G}}^0_{\cal{I}}$ we only need to generate a pseudo-random number from $\Gamma(L,L)$ and then divide it by a pseudo-random number obtained from $\Gamma(-\alpha, 1/\gamma)$. For instance, the following {\tt Ox} \cite{Doornik2002} function can be used for ${\cal{G}}^0_{\cal{I}}$ random number generation:
{\scriptsize
\begin{verbatim}
// A - texture parameter
// G - scale parameter
g0i_generation (const A, const G)
{
    decl x = rangamma(rowss, cols, -A, 1/G);
    decl y = rangamma(rowss, cols, L, L);
    return y./x;
}
\end{verbatim}
}

In all simulations, $L=1$. Recall that this is the most challenging situation. All images we generated are rectangular and contain $20\times 100$ pixels ($N=100$) with an edge located at $j=50$. Thus, each image contains two distinct regions of equal size. In the region to the left of the edge $\alpha = \alpha_\ell$, and otherwise $\alpha = \alpha_r$, with $\alpha_\ell, \alpha_r \in \{-2, -3, \ldots, -15\}$. For each region, the scale parameter $\gamma$ was chosen according to (\ref{eq:gamma.norm}). The most challenging situation as far as edge detection is concerned takes place when $|\alpha_r - \alpha_\ell| = 1$; the higher the absolute difference between $\alpha_r$ and $\alpha_\ell$, the easier it is to locate the edge. We notice that the case in which $|\alpha_r - \alpha_\ell| = 1$ becomes particularly more challenging when the values $\alpha_r$ and $\alpha_\ell$ are large (in absolute values) because we are then dealing with two very homogeneous regions. Accordingly, \cite{Nascimento2010} shows that when $\alpha_\ell,\alpha_r\rightarrow-\infty$ the stochastic distance between the two ${\mathcal{G}}^0_{\mathcal{I}}$ diminishes, even when $|\alpha_r - \alpha_\ell|$ remains constant.

The number of Monte Carlo replications is $R=5000$ and the number of bootstrap replications is $B = 1000.$ For each configuration, we constructed $5000$ confidence intervals using the basic bootstrap method, the percentile method, and also the ST1 and ST2 methods. The nominal coverage is $1-a=0.95$ $(95\%)$. The results for the ST1 and ST2 methods were obtained using $B'=50,$ and $B''=200$, and for ST2 we used $B_{x} = 200$.

\begin{figure}[!h]
  \begin{center}
    \ifthenelse {\boolean{fig_pdf}}
  	{
   		\includegraphics*[width=9.3cm, viewport= 18 18 594 774]{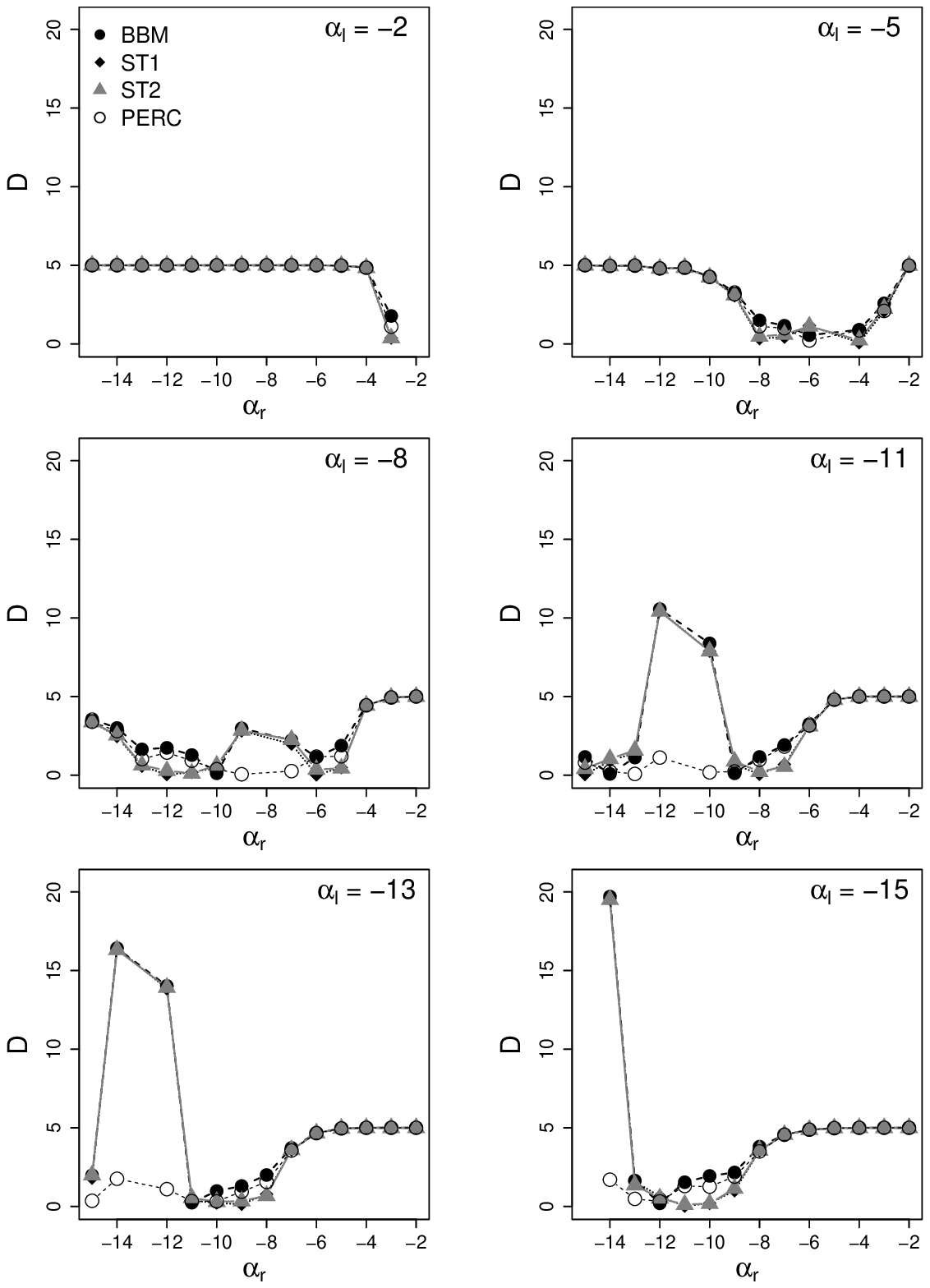}
    }
  	{
   		\includegraphics*[height=18cm, viewport= 18 18 594 774]{graficosCovering.ps}
    }
    \caption{Distances between empirical and nominal coverages.} 
    \label{fig.selected.covering}
  \end{center}
\end{figure}

We computed the exact coverage ($\bar{C}$) of each interval estimator corresponding to the 95\% nominal level, which was done for each configuration $(\alpha_\ell, \alpha_r)$. The best performing estimator is the one whose exact coverage is closest to the nominal coverage. Figure~\ref{fig.selected.covering} plots the distance $D = |\bar{C} - 0.95|\times100\%$ against $\alpha_r$ (the roughness parameter value to the right of the edge). The six panels in the figure correspond to $\alpha_\ell \in \{-2,-5,-8,-11,-13,-15\}$. Situations in which $\alpha_\ell=\alpha_r$ (i.e., there is no edge) are not considered.

Visual inspection of Figure~\ref{fig.selected.covering} reveals that all interval estimators perform equally well when $\alpha_\ell=-2$ and $\alpha_{r}\leq-4$; their coverages are 100\%. We also notice that as   $|\alpha_\ell-\alpha_{r}|$ increases the differences between the different coverages tend to decrease, i.e., the different intervals display similar coverages. That happens because the distribution of $\hat{\jmath}_{KW}$ tends to be more concentrated around of the true value $j=50$ as  
 $|\alpha_\ell-\alpha_{r}|$ increases, i.e., the estimator $\hat{\jmath}_{KW}$ is more accurate when the two regions have very distinct textures. Such a tendency is not, however, uniform. Also, 
when the regions on both sides of the edge are very heterogeneous (for example, $\alpha_\ell=-2,$ and $\alpha_r=-3,$ or $\alpha_\ell=-5$ and $\alpha_\ell=-4$), the different intervals behave similarly, but when the two regions of are very homogeneous (for example $\alpha_\ell=-13,$ and $\alpha_r=-12$), the percentile estimator outperforms all other interval estimators. Figure~\ref{fig.selected.covering} shows that the percentile estimator is the best performing interval estimator in most situations. It is also noteworthy that the percentile estimator is the best performing estimator in the most challenging situations, i.e., whenever the two regions are very homogeneous and have similar textures. 

\begin{figure}[!h]
\centering \hspace{1cm} ${\otimes}$ {\scriptsize BBM,  $\bigcirc$ PERC, $\blacklozenge$ ST1,  $\blacktriangle$ ST2, $\boxtimes$ ALL METHODS }  \vspace{0.1cm}
  \begin{center}
    \ifthenelse {\boolean{fig_pdf}}
  	{
   		\includegraphics*[width=8.5cm, viewport= 63 159 543 590]{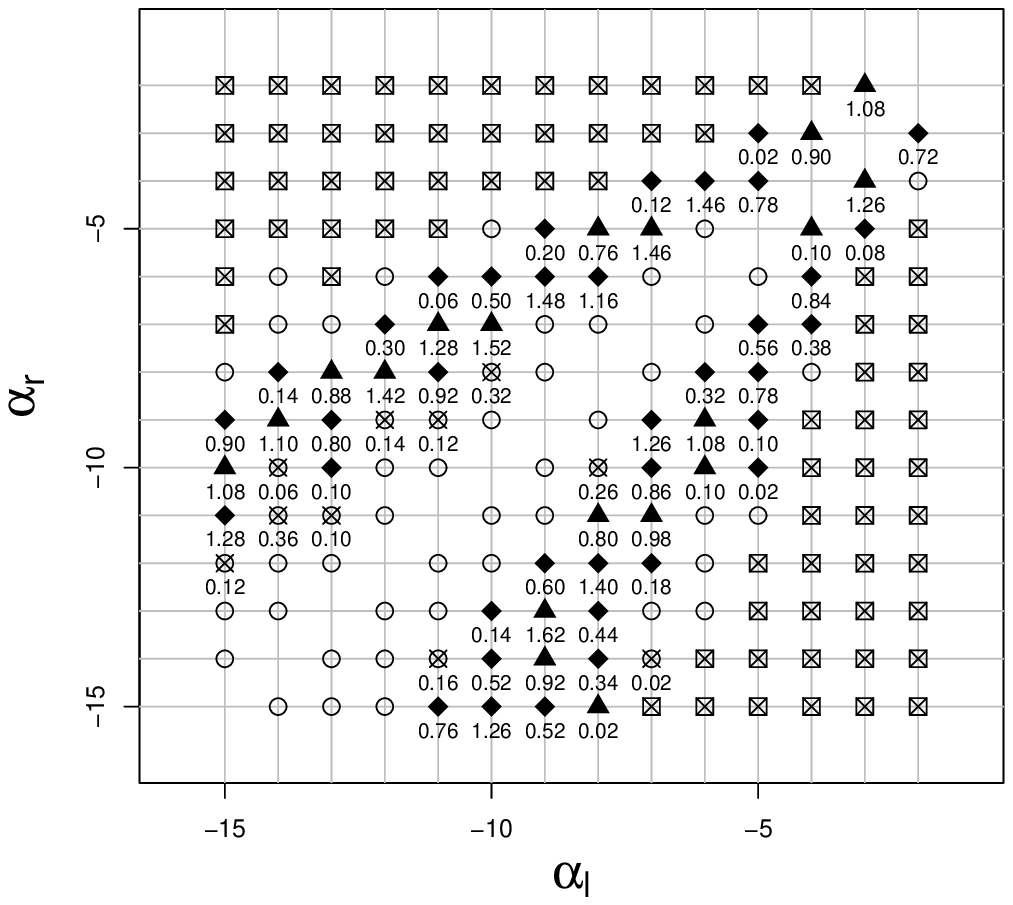}
    }
  {
   		\includegraphics*[height=12cm, viewport= 67 159 543 590]{todasCoberturasDiffPerc.ps}
    }
    \caption{Best method and difference performance $\Delta$ between the percentile and the best performer.}
    \label{fig.best.covering}
  \end{center}
\end{figure}

In Figure~\ref{fig.best.covering}, a symbol indicates which is the best performing interval estimator for each configuration $(\alpha_\ell, \alpha_r)$; the quantity bellow the symbol is 
$$
\Delta = D_{\mbox{percentile}} - D_{\mbox{best method}},
$$
which measures the method performance relative to the percentile method. The symbols used in the figure are the same as the ones used in Figure~\ref{fig.selected.covering}, except for the symbol used to represent BBM, which is now $\otimes$. The symbol $\boxtimes$ is used whenever all methods produced the same coverage. We observe that in configurations in which the two absolute values of roughness parameter are large and similar (the most challenging situations), the percentile estimator is the best performer, closely followed by BBM. 

The two studentized estimators, ST1 and ST2, outperformed the competition whenever $|\alpha_\ell-\alpha_{r}| \leq 5$. In these situations, the values of $\Delta$ are between $1\%$ and $2\%$. Since these two estimators behave similarly, we recommend the use of ST1, which is less computationally intensive. It should be noted that in such situations the percentile estimator is quite competitive with the studentized estimators. 

All methods behave similarly when the absolute difference between $\alpha_\ell$ and $\alpha_r$ becomes very large; see Figure~\ref{fig.selected.covering}. 

When taken together, our numerical results show that the percentile estimator is either the best performing estimator or quite competitive with the best performer. Therefore, we conclude that the percentile edge estimator is to be preferred.

\begin{figure}[!h]
  \begin{center}   
    \ifthenelse {\boolean{fig_pdf}}
  	{
    	\ifthenelse {\boolean{fig_eng}}
    	{
    		\includegraphics[width=7cm]{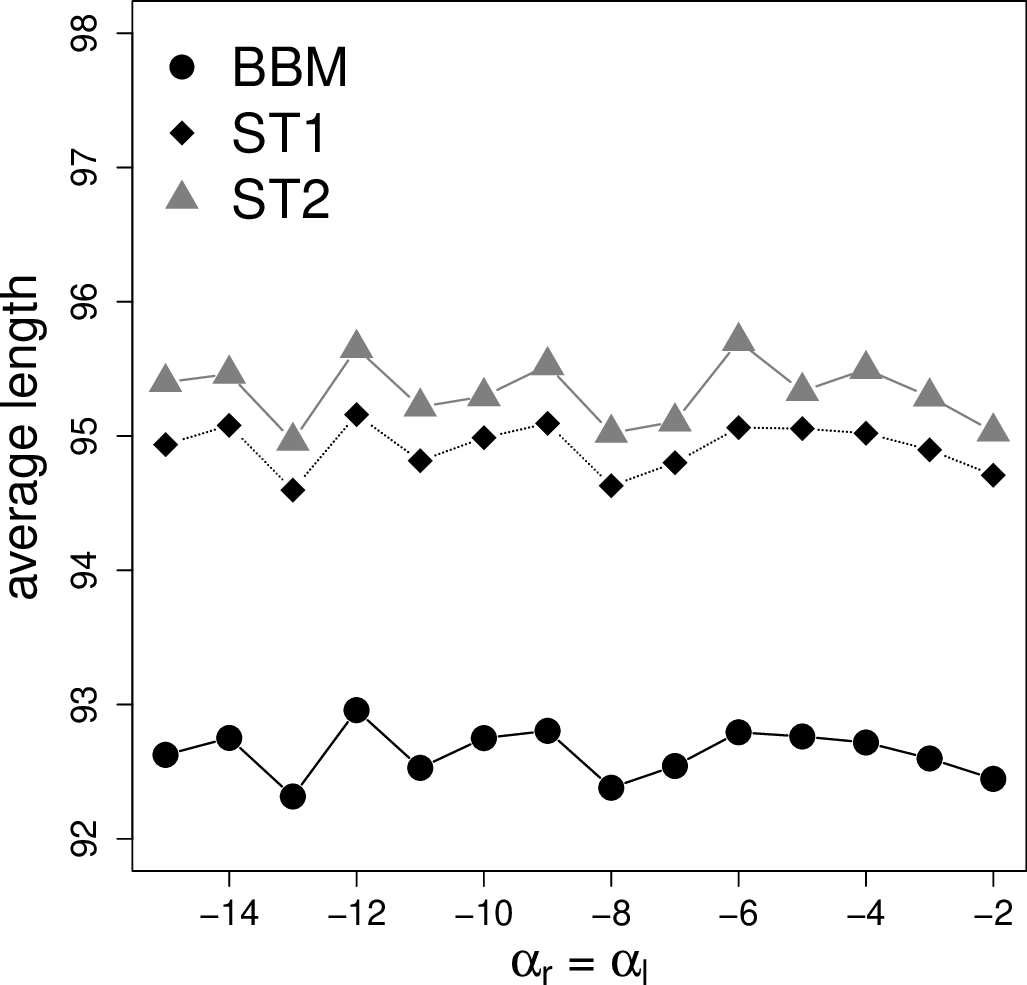}
    	}
    	{
    		\includegraphics[height=7cm]{semBorda-Amplitude-pt.eps}
    	}
    }
    {
    	\ifthenelse {\boolean{fig_eng}}
    	{
    		\includegraphics[height=7cm]{semBorda-Amplitude.eps}
    	}
    	{
    		\includegraphics[height=7cm]{semBorda-Amplitude-pt.eps}
    	}
    }
    \caption{Average length for situations where there is not edge.}
    \label{fig.semborda-amplitude}
  \end{center}
\end{figure}

We have also carried out simulations in which there is no edge, i.e., $\alpha_\ell = \alpha_r$. Here, we shall focus not on coverages but on the average lengths of the different confidence intervals. The estimated edge locations are expected to be randomly and uniformly distributed along the detection line since there is no edge. We thus expect the average interval length to be approximately equal to the detection line length.  Figure~\ref{fig.semborda-amplitude} displays average interval lengths (AIL).

We notice that all methods produce similar average interval lengths, regardless of the configuration $(\alpha_\ell, \alpha_r)$, as expected. The percentile and BBM interval estimators have the same average length: approximately $93$ pixels. Notice that an open circle is used to represent the percentile method
and a filled circle is used for the BBM method, therefore, only the filled circle (BBM) is visible.
The ST1 and ST2 estimators displayed similar average lengths (approximately $95$ pixels), the former always being slightly narrower. Overall, we notice that all methods yield very wide intervals when there is no edge to be located. A very wide confidence interval can then be taken as evidence that there is no edge. 

Table~\ref{tabela.mean.range} displays selected average interval lengths. Notice that when $|\alpha_\ell-\alpha_{r}|$ is large, the average lengths are null or very small. On the other hand, when $|\alpha_\ell-\alpha_{r}|$ is small, the average lengths tend to be large. percentile and BBM intervals yields the smallest average lengths whenever the average lengths are not all the same. The largest average lengths correspond to the studentized interval estimators, ST1 and ST2. The average lengths of the studentized and percentile intervals are very close.

 \setlength{\tabcolsep}{3pt}
\begin{table}[!h]
\centering
\caption{Average interval lengths (AIL) for $\alpha_\ell\neq\alpha_r$ (selected configurations).}
\label{tabela.mean.range}
{\scriptsize
\renewcommand{\arraystretch}{1.1} 
\begin{tabular}{@{}c|ccc|ccc|ccc}
  \hline
 & & $\alpha_\ell = -3$ & & & $\alpha_\ell = -8$ &  & & $\alpha_\ell = -13$ & \\
  \hline
$\alpha_r$ & BBM & ST1 & ST2 & BBM & ST1 & ST2 & BBM & ST1 & ST2 \\ 
  \hline
 -2 & 1.04 & 1.07 & 1.10 & 0.00 & 0.00 & 0.00    & 0.00 & 0.00 & 0.00    \\ 
 -3 &      &      &      & 0.00 & 0.00 & 0.00    & 0.00 & 0.00 & 0.00    \\ 
 -4 & 3.40 & 3.77 & 3.84 & 0.05 & 0.04 & 0.04    & 0.00 & 0.00 & 0.00    \\ 
 -5 & 0.50 & 0.47 & 0.49 & 0.99 & 1.02 & 1.05    & 0.00 & 0.00 & 0.00    \\ 
 -6 & 0.04 & 0.03 & 0.03 & 3.96 & 4.35 & 4.40    & 0.02 & 0.02 & 0.02    \\ 
 -7 & 0.01 & 0.00 & 0.00 & 24.39 & 25.90 & 26.16 & 0.20 & 0.18 & 0.19    \\ 
 -8 & 0.00 & 0.00 & 0.00 &       &       &       & 0.96 & 0.99 & 1.03    \\ 
 -9 & 0.00 & 0.00 & 0.00 & 32.57 & 34.23 & 34.65 & 2.29 & 2.59 & 2.63    \\ 
-10 & 0.00 & 0.00 & 0.00 & 7.28 & 7.89 & 8.00    & 5.09 & 5.54 & 5.62    \\ 
-11 & 0.00 & 0.00 & 0.00 & 3.21 & 3.55 & 3.60    & 14.59 & 15.72 & 15.91 \\ 
-12 & 0.00 & 0.00 & 0.00 & 1.79 & 2.00 & 2.05    & 62.06 & 63.61 & 64.14 \\ 
-13 & 0.00 & 0.00 & 0.00 & 0.97 & 1.00 & 1.02    &       &       &       \\ 
-14 & 0.00 & 0.00 & 0.00 & 0.45 & 0.43 & 0.45    & 67.00 & 68.57 & 69.14 \\ 
-15 & 0.00 & 0.00 & 0.00 & 0.19 & 0.17 & 0.18    & 21.58 & 23.01 & 23.34 \\
   \hline
\end{tabular}}
\end{table}
 \setlength{\tabcolsep}{6pt}

We have also considered the case in which the edge lies not in the middle of the detection window. In particular, we carried out simulations using $j=20$. The number of looks ($L=1$) and the detection window size ($20\times100$ pixels) were kept constant. The results corresponding to low contrast situations, which are the most challenging cases, are presented in Table~\ref{tabela.borda.nao.central}. Such results are in agreement with those previously reported; see  Table~\ref{tabela.mean.range} and Figure~\ref{fig.selected.covering}.

\begin{table}[!h]
\centering
\caption{Edge lies not in the middle of the detection window -- Percentile Method (selected configurations).}
\label{tabela.borda.nao.central}
{\scriptsize
\renewcommand{\arraystretch}{1.1} 
\begin{tabular}{@{}c|c|c|c|c}
  \hline
  \hline
Texture & $\alpha_\ell$ & $\alpha_r$ & AIL & coverage (\%) \\ 
  \hline
\multirow{2}{*}{Extremely Heterogeneous}  & -2  &  -3 &  1.18 & 96.40 \\
 & -2  &  -4 &  0.04 & 99.68 \\
 \hline
\multirow{2}{*}{Heterogeneous}  & -7  &  -8 & 33.70 & 94.74 \\
 & -7  &  -9 &  6.00 & 94.90 \\
 \hline
\multirow{2}{*}{Homogeneous}  & -13 & -15 & 29.40 & 94.16 \\ 
 & -14 & -15 & 78.76 & 96.22 \\ 
   \hline
   \hline
\end{tabular}}
\end{table}
 
The execution times of the different interval edge estimators for each Monte Carlo replication are given in Table~\ref{tabela.custo}. Since the computational cost does not depend on the parameter configuration, we use $(\alpha_\ell, \alpha_r) = (-2,-3)$ in order to measure the different execution times. We report both the total time (in seconds) it takes to compute the interval estimator and that relative to the bootstrap-$t$ estimator (\%). The bootstrap-$t$ estimator was computed using 50 replications in the inner bootstrap, which is carried out for variance estimation. The results show that the Bootstrap-$t$ is approximately 50 times more computationally intensive than all competing methods. Percentile and BBM methods have nearly the same computational cost. ST1 and ST2 are slightly more costly, ST1 being less computationally intensive than ST2. It should be noted that the percentile method is not only the best performer as far as coverage is concerned, but it is also the least computationally intensive method.

\begin{table}[!h]
\begin{center}
\caption{Computational cost and computational cost relative to bootstrap-$t$.}
\label{tabela.custo}
{\scriptsize
\renewcommand{\arraystretch}{1.1} 
\begin{tabular}{@{} lrr}
\hline\hline
\textbf{Method} & \textbf{Execution time (s)} & \textbf{Relative to bootstrap-$t$ $(\%)$}\\
\hline
Bootstrap-$t$ & $81.422$ & $100.00$ \\
BBM           & $ 1.578$ & $1.94  $ \\
ST1           & $ 1.630$ & $2.00  $ \\
ST2           & $ 2.191$ & $2.69  $ \\
PERC          & $ 1.565$ & $1.92  $ \\
\hline\hline
\end{tabular} }

\end{center}
\end{table}

The sample size affects the point estimates accuracy. Interval estimates constructed using smaller samples tend to be wider, thus reflecting the increased uncertainty. We observed such a behavior in Monte Carlo simulations (results not included here for brevity).

\section{An application}
\label{S:application}

\begin{figure}[!h]
\begin{center}
\ifthenelse {\boolean{fig_pdf}}
{
	\includegraphics*[width=8.5cm, viewport= 0 569 595 842]{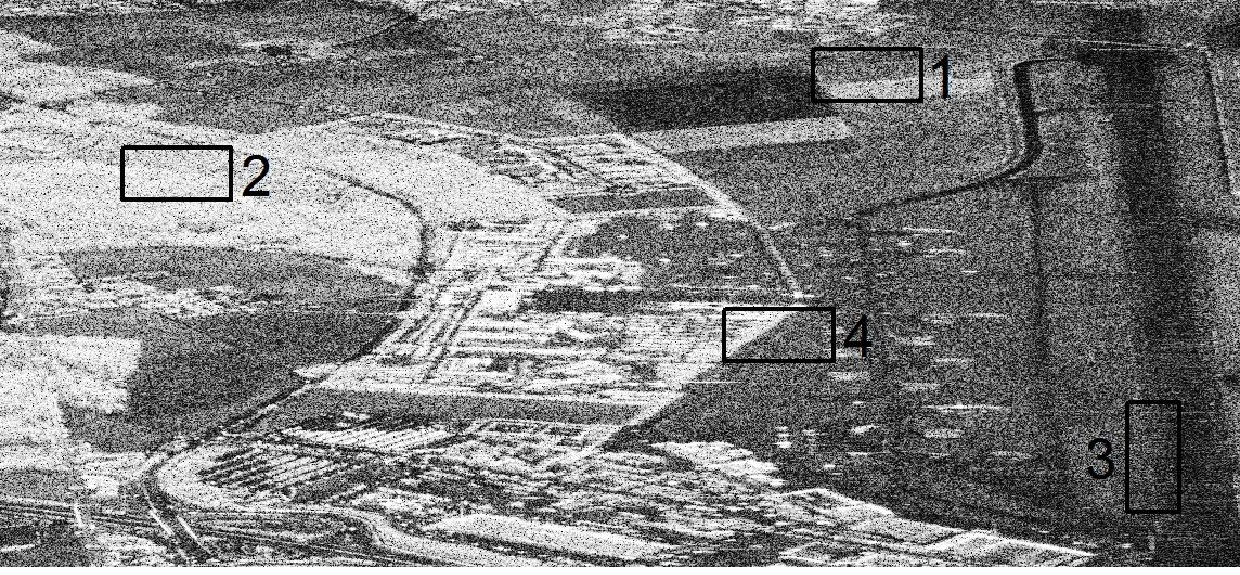}
}
{
	\includegraphics[width=13.65cm]{regioesEmEscala.eps}
}
\caption{E-SAR image with $1100\times2400$ pixels, and delimited regions.}%
\label{fig.imagemSAR}
\end{center}
\end{figure}

In this section we apply the different edge interval estimators to a real image, i.e., to observed (not simulated) data. An E-SAR single look image of We\ss{} ling (Bayern, Germany) was used \cite{Horn1998}. This image is displayed in Figure~\ref{fig.imagemSAR}.

\noindent It was obtained in L band, and it exhibits part of an airport, urban areas, and pastures. The rectangles in the image are the regions chosen for edge detection. Rectangles (regions) $1,2$ and $4$ have $210 \times 100$ pixels whereas rectangle (region) $3$ has $100 \times 210$ pixels. In the edge detection process, we divided rectangle $3$ in ten horizontal windows, each of $21\times100$ pixels, the detection line being horizontal for each window and dividing the window in equal parts of $10\times100$ pixels. Similarly, each of the other rectangles was divided in ten windows of $100\times21$ pixels, the detection line being vertical and dividing the window in equal parts of $100\times10$ pixels. Therefore, for each rectangle we compute ten interval edge estimates. In the figures that follow, we use green lines to connect the lower interval limits and also the upper interval limits; red lines are used to connect the different edge point estimates. Each detection window is delimited by yellow lines, the windows being numbered from left to right or from top to bottom (for example, the first  window in rectangle $1$ corresponds to the window located in the extreme left).

\begin{figure}[!h]
\begin{center}
\subfloat[fig.regiao1a][]{\includegraphics*[width=7cm]{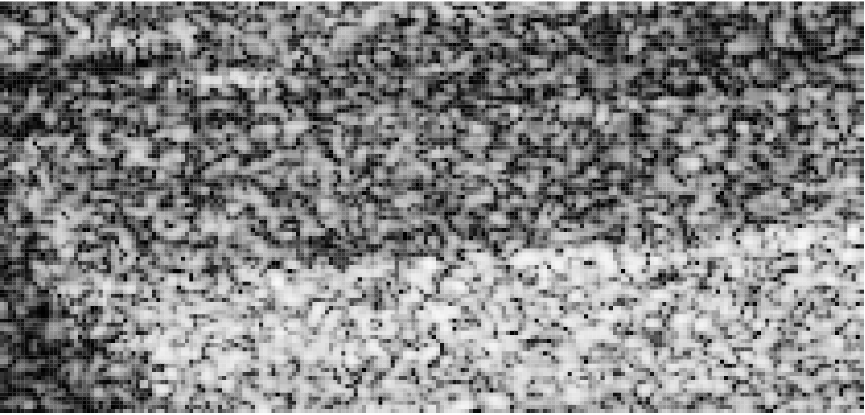}}\\
\subfloat[fig.regiao1b][]{\includegraphics*[width=7cm, bb= 104 328 520 528]{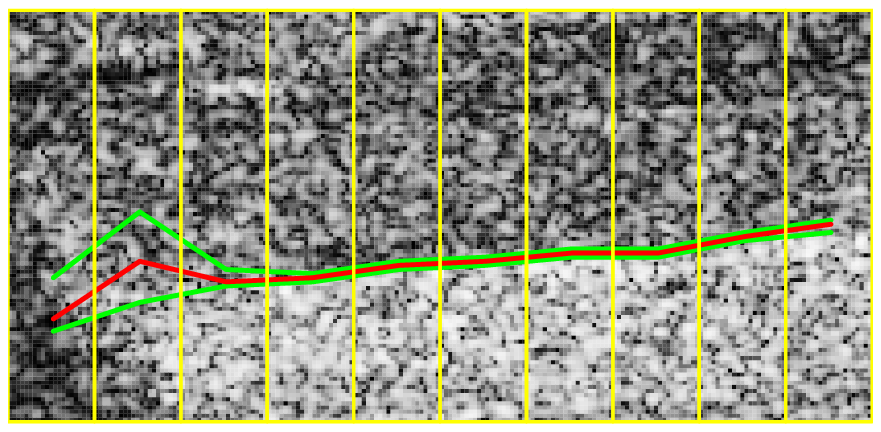}}
\caption{Results for interval detection in region 1. (a) Region 1 magnification. (b) Graphical presentation of confidence intervals to edges in region 1. Green lines links interval limits for each detection window. The windows are drawn by yellow lines. The red line links edge point estimates for each detection window.}%
\label{fig.regiao1}%
\end{center}
\end{figure}

Figure~\ref{fig.regiao1} contains results relative to edge detection in rectangle $1$. 

\noindent We observe that the confidence intervals are quite narrow (the smallest interval length equals $3$ pixels), except for the first two detection windows. In the second window there are less pixels in light regions than in dark regions, in which there is more noise, hence the wider confidence interval. This happens because edge detection becomes quite challenging under complex, rich textures. We also note that the estimated edge is typically located in areas in which the differences in texture are largest; see, e.g., the first window.  

\begin{figure}[!h]
\begin{center}
\subfloat[fig.regiao2a][]{\includegraphics*[width=7cm, bb= 104 328 520 528]{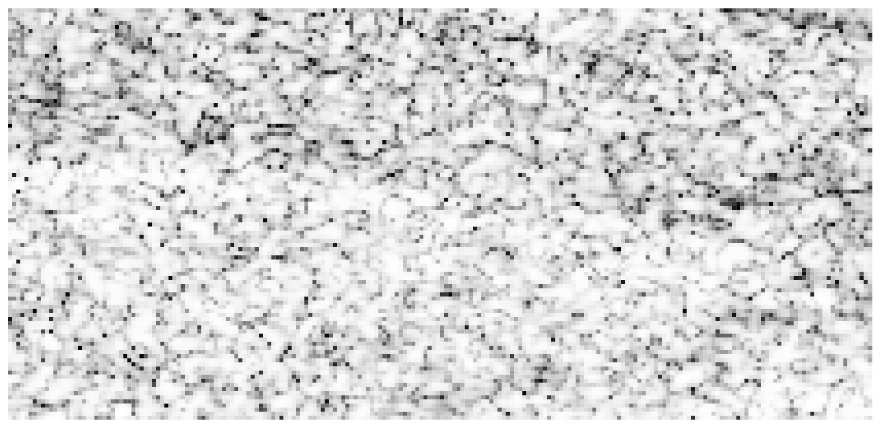}}\\
\subfloat[fig.regiao2b][]{\includegraphics*[width=7cm, bb= 104 328 520 528]{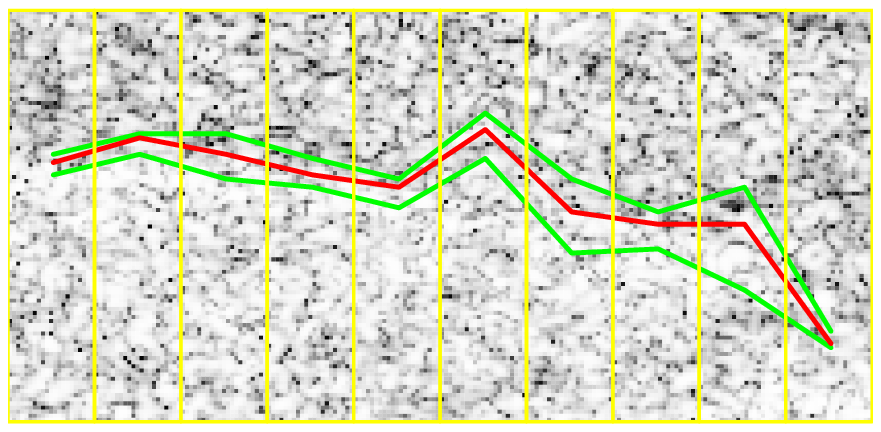}}
\caption{Results for interval detection in region 2. (a) Region 2 magnification. (b) Graphical presentation of confidence intervals to edges in region 2. Green lines links interval limits for each detection window. The windows are drawn by yellow lines. The red line links edge point estimates for each detection window.}%
\label{fig.regiao2}%
\end{center}
\end{figure}

A challenging situation for edge detection can be seen in rectangle (region) $2$ of Figure~\ref{fig.imagemSAR}. There we notice two fairly different textures, but it is not easy to identify where exactly lies the frontier. The results of edge detection for this region are presents in Figure~\ref{fig.regiao2}. Note that, in all detection windows, the confidence intervals are fairly wide; the smallest interval length is $5$ pixels (last detection window) and the widest interval covers $26$ pixels (next to last detection window). In challenging situations such as this interval estimation becomes quite useful since it signals that point edge detection may not be reliable.  

\begin{figure}[!h]
\begin{center}
\subfloat[fig.regiao3a][]{\includegraphics*[height=7cm, bb= 222 242 402 614]{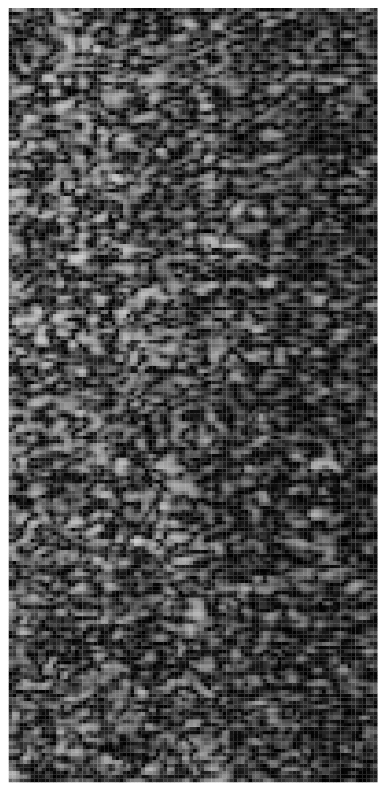}}
\qquad
\subfloat[fig.regiao3b][]{\includegraphics*[height=7cm, bb= 222 242 402 614]{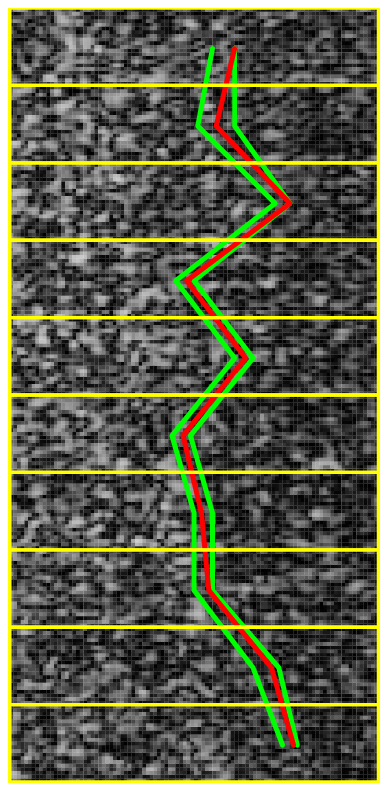}}
\caption{Results for interval detection in region 3. (a) Region 3 magnification. (b) Graphical presentation of confidence intervals to edges in region $3$. Green lines links interval limits for each detection window. The windows are drawn by yellow lines. The red line links edge point estimates for each detection window.}%
\label{fig.regiao3}%
\end{center}
\end{figure}

Edge detection in rectangle (region) $3$ of Figure~\ref{fig.imagemSAR} is even more challenging than in rectangle (region) $2$  because the edge divides the image in two regions with textures that are visually very diffuse. The detection results are displayed in Figure~\ref{fig.regiao3}. 

\noindent We note that the interval lengths are nearly the same in all detection windows. None of interval lengths exceeds $11$ pixels. We conclude that, despite the aforementioned difficulty, the KW detector yields accurate edge estimates.

\begin{figure}[!h]
\begin{center}
\subfloat[fig.regiao4a][]{\includegraphics*[width=7cm, bb= 104 328 520 528]{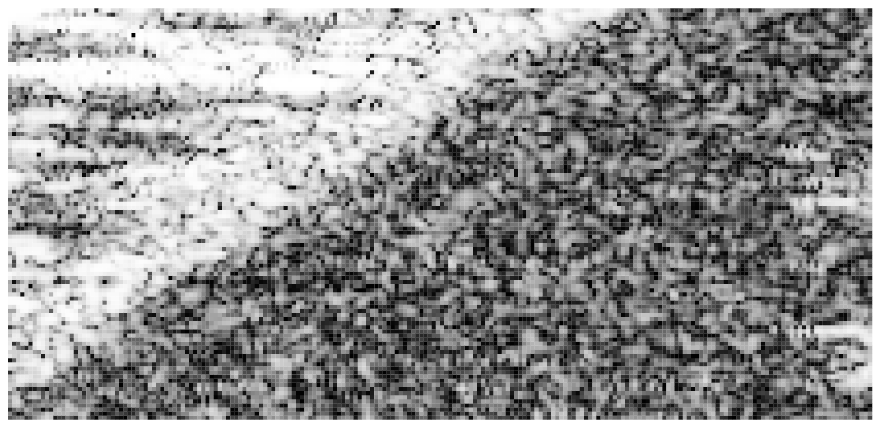}}\\
\subfloat[fig.regiao2b][]{\includegraphics*[width=7cm, bb= 104 328 520 528]{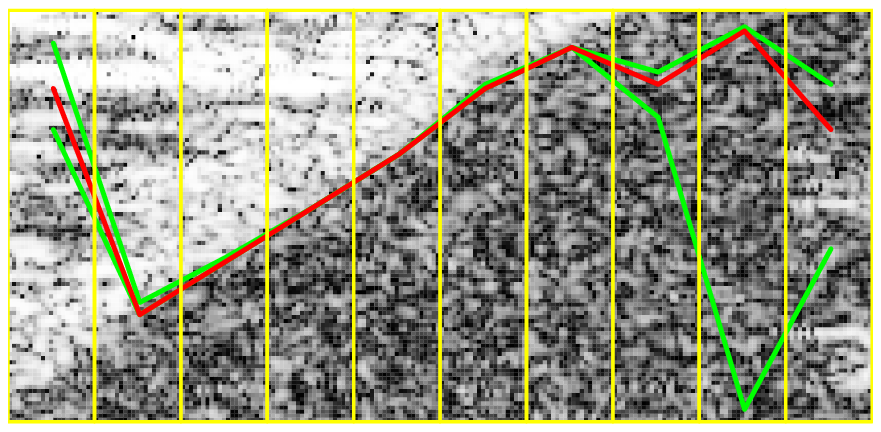}}
\caption{Results for interval detection in region 4. (a) Region 4 magnification. (b) Graphical presentation of confidence intervals to edges in region 4. Green lines links interval limits for each detection window. The windows are drawn by yellow lines. The red line links edge point estimates for each detection window.}%
\label{fig.regiao4}%
\end{center}
\end{figure}

A limitation of the point edge detection methods we consider is that an edge will be always located even when there is none. This happens because $T_{KW}$ in (\ref{eq:funcaoObjetivo.kw}) always achieves a maximum value. Interval estimation can, however, be used to assess whether an edge indeed exists. Such a situation is explored in rectangle (region) $4$ of Figure~\ref{fig.imagemSAR}. A confidence interval that covers $94$ pixels is obtained, thus suggesting that there is no edge. The results of edge detection for this region are presents in Figure~\ref{fig.regiao4}. In the last window detection, we obtain a fairly wide interval ($41$ pixels) which can be taken as evidence that the point edge estimate is not accurate. A similar situation takes place at the first detection window. Here, however, the dark regions are most likely not the result of noise but reflect the terrain characteristics.

Samples $2$ and $3$ are particularly interesting since they involve neighboring regions that are very similar. The edge is thus difficult to be located and the lengths of the corresponding interval estimates signal such uncertainty. They are wider in areas where the contrast between the two regions is small.

\begin{figure}[!h]
\begin{center}
\subfloat[fig.suavizacaoa][Gaussian kernel smoothing with bandwidth equal to 23.]{\includegraphics*[width=7cm]{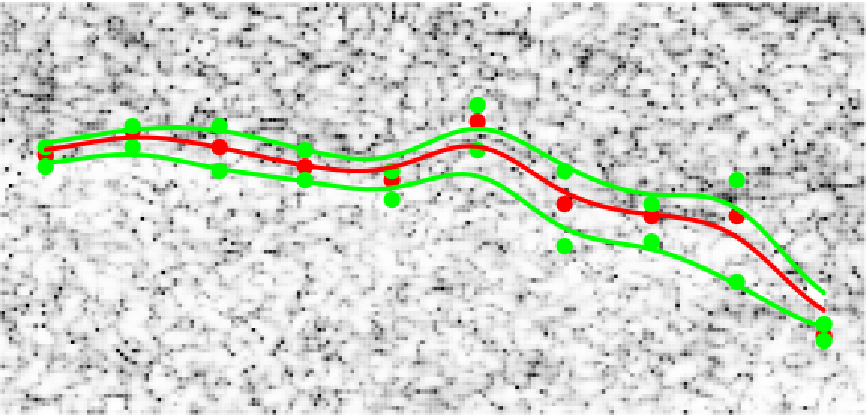}}\\
\subfloat[fig.suavizacaob][Cubic Splines smoothing with degrees of freedom equal to 8.]{\includegraphics*[width=7cm]{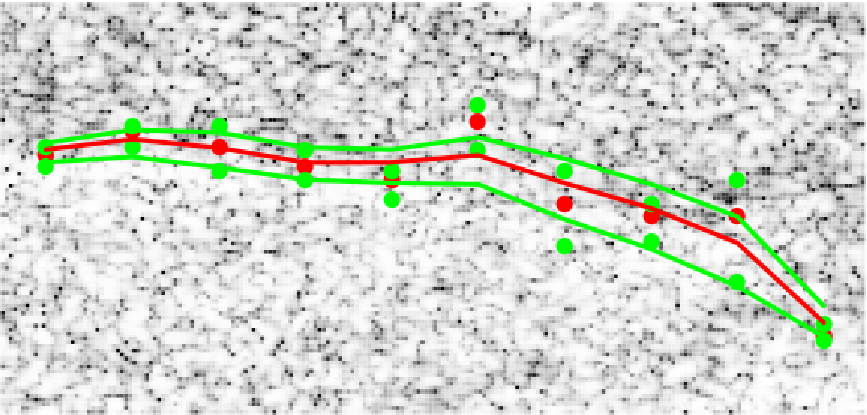}}
\caption{Smooth confidence bands, region $2$. Point estimates are denoted by red dots and the different confidence intervals are marked using green dots.}%
\label{fig.suavizacao}%
\end{center}
\end{figure}

In Figures~\ref{fig.regiao1}, \ref{fig.regiao2}, \ref{fig.regiao3} and \ref{fig.regiao4} we connected the different confidence interval limits using straight lines. As noted by a referee, when using interval edge detection, the user can specify the kind of edge she is willing to retrieve. For instance, provided a wide enough interval, she can choose between linear features and straight angles between them, as is the case of many agricultural fields, or smoothly varying edges, as in shores. This is, we believe, a novel feature of our approach. In particular, we note that smoothing methods, such as kernel and splines smoothing \cite{suavizacao1,suavizacao2,suavizacao3}, can be used for constructing the lower and upper curves. This was done in Figure~\ref{fig.suavizacao}. We note that the user can choose the amount of smoothing she wants to use.

\begin{figure*}[!H]
\begin{center}

  \subfloat[\scriptsize{Extremely Heterogeneous Regions.}]{\label{fig:1}
    \begin{pspicture}(0,0)(2.8,5.6)
      \rput(1.4,2.8){\includegraphics[scale=0.4, angle=90]{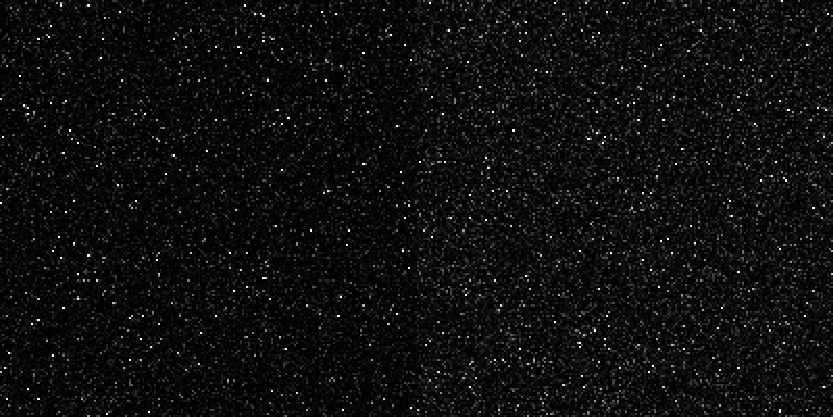}}
      \rput[bc](1.4,4.2){\huge {\white $\alpha=-3$}}
      \rput[bc](1.4,1.4){\huge {\white $\alpha=-2$}}
    \end{pspicture}}\qquad
  \subfloat[\scriptsize{Edge confidence interval for (a).}]{\label{fig:2}
    \begin{pspicture}(0,0)(2.8,5.6)
      \rput(1.4,2.8){\includegraphics[scale=0.4, angle=90]{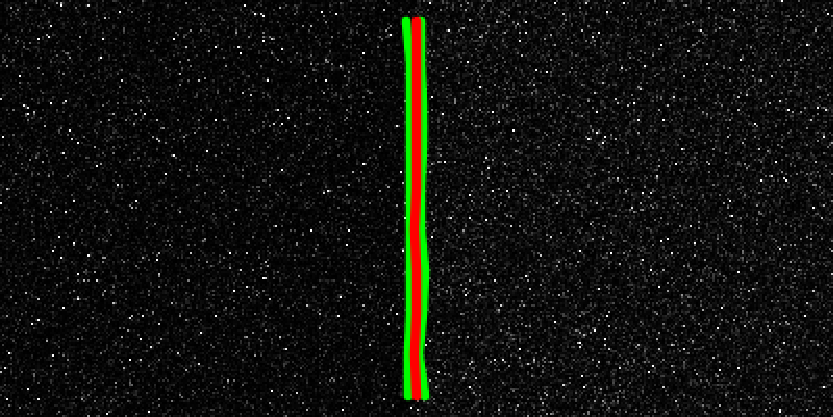}}
    \end{pspicture}}\qquad
  \subfloat[\scriptsize{{\it gPb} method applied to (a).}]{\label{fig:3}
    \begin{pspicture}(0,0)(2.8,5.6)
      \rput(1.4,2.8){\includegraphics[scale=0.4, angle=90]{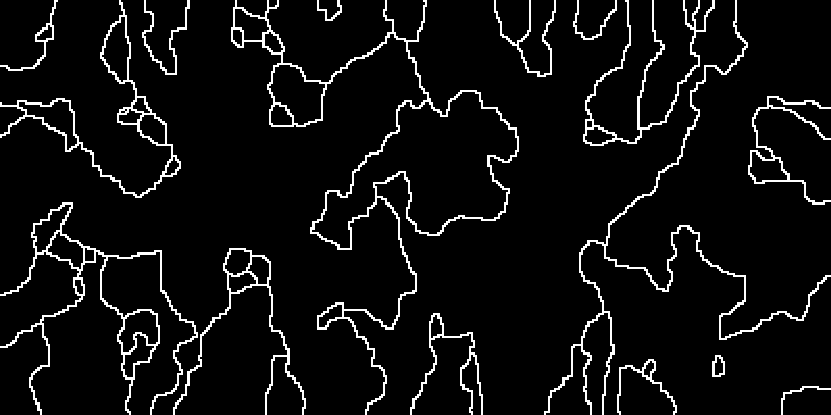}}
    \end{pspicture}}\qquad
  \subfloat[\scriptsize{Sobel method applied to (a).}]{\label{fig:4}
    \begin{pspicture}(0,0)(2.8,5.6)
      \rput(1.4,2.8){\includegraphics[scale=0.4, angle=90]{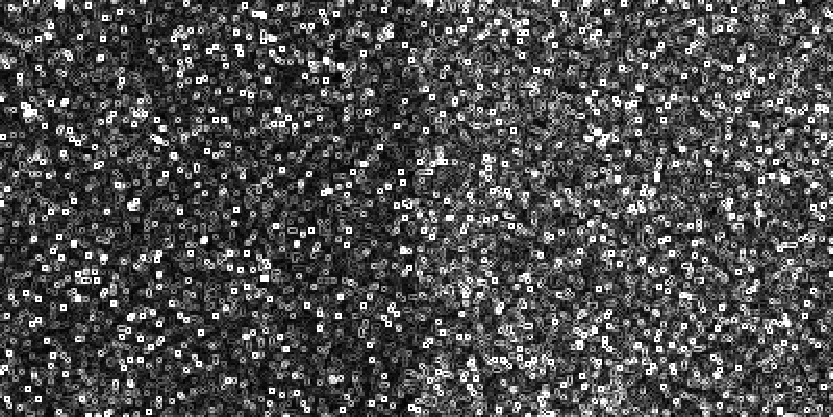}}
    \end{pspicture}}\\

  \subfloat[\scriptsize{Heterogeneous Regions.}]{\label{fig:5}
    \begin{pspicture}(0,0)(2.8,5.6)
      \rput(1.4,2.8){\includegraphics[scale=0.4, angle=90]{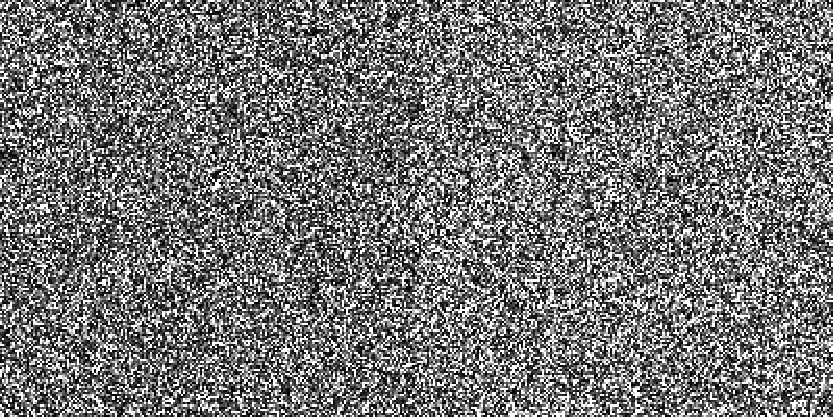}}
      \rput[bc](1.4,4.2){\huge{\textcolor{white}{$\alpha=-8$}}}
      \rput[bc](1.4,1.4){\huge{\textcolor{white}{$\alpha=-7$}}}
    \end{pspicture}}\qquad
  \subfloat[\scriptsize{Edge confidence interval for (e).}]{\label{fig:6}
    \begin{pspicture}(0,0)(2.8,5.6)
      \rput(1.4,2.8){\includegraphics[scale=0.4, angle=90]{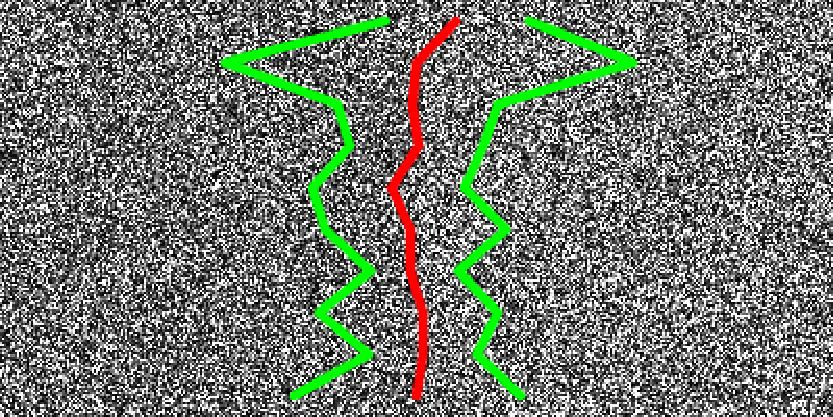}}
    \end{pspicture}}\qquad
  \subfloat[\scriptsize{{\it gPb} method applied to (e).}]{\label{fig:7}
    \begin{pspicture}(0,0)(2.8,5.6)
      \rput(1.4,2.8){\includegraphics[scale=0.4, angle=90]{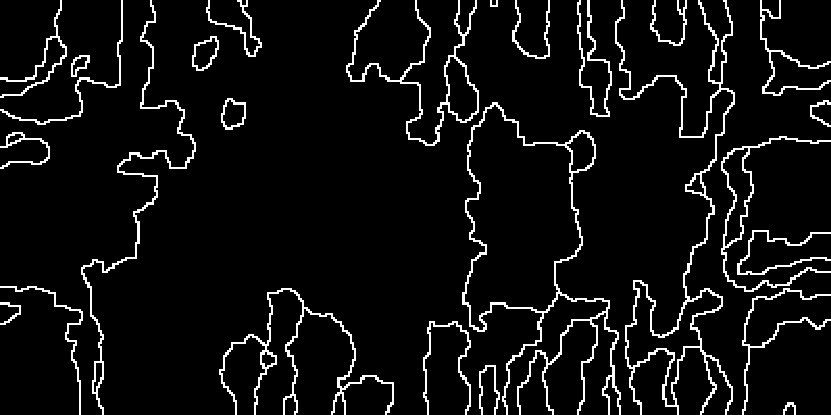}}
    \end{pspicture}}\qquad
  \subfloat[\scriptsize{Sobel method applied to (e).}]{\label{fig:8}
    \begin{pspicture}(0,0)(2.8,5.6)
      \rput(1.4,2.8){\includegraphics[scale=0.4, angle=90]{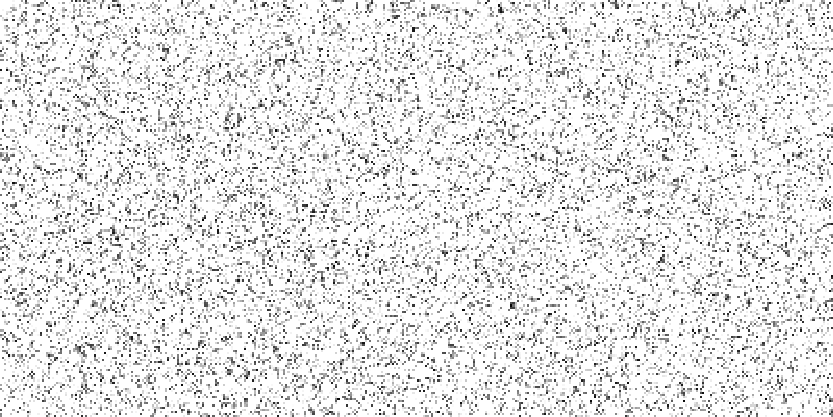}}
    \end{pspicture}}\\    

  \subfloat[\scriptsize{Homogeneous Regions.}]{\label{fig:9}
    \begin{pspicture}(0,0)(2.8,5.6)
      \rput(1.4,2.8){\includegraphics[scale=0.4, angle=90]{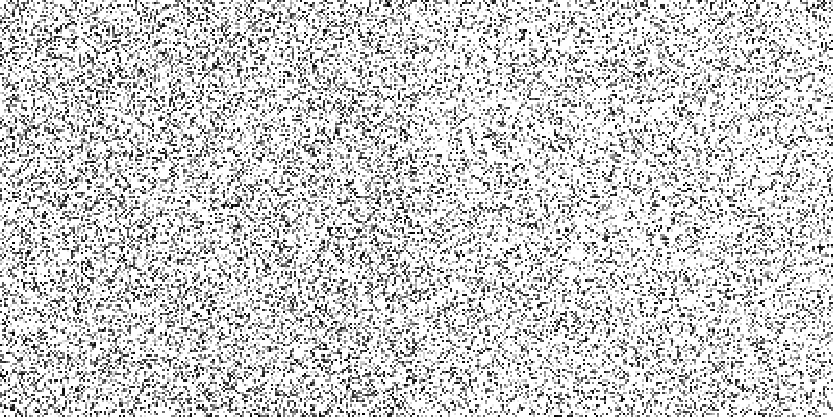}}
      \rput[bc](1.4,4.2){\huge{\textcolor{black}{$\alpha=-15$}}}
      \rput[bc](1.4,1.4){\huge{\textcolor{black}{$\alpha=-14$}}}
    \end{pspicture}}\qquad
  \subfloat[\scriptsize{Edge confidence interval for (i).}]{\label{fig:10}
    \begin{pspicture}(0,0)(2.8,5.6)
      \rput(1.4,2.8){\includegraphics[scale=0.4, angle=90]{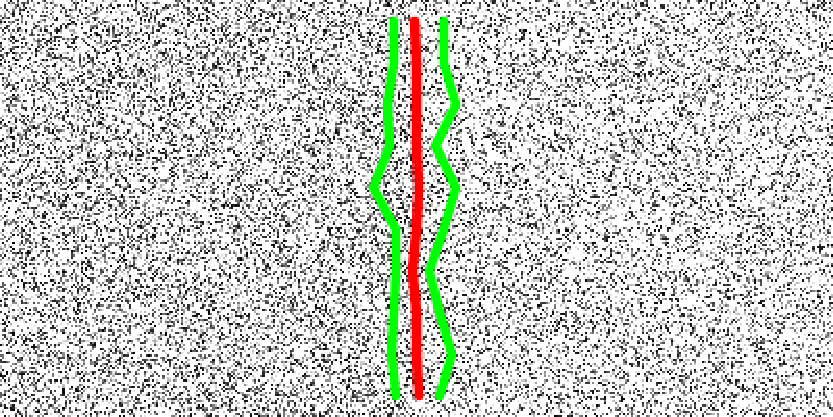}}
    \end{pspicture}}\qquad
  \subfloat[\scriptsize{{\it gPb} method applied to (i).}]{\label{fig:11}
    \begin{pspicture}(0,0)(2.8,5.6)
      \rput(1.4,2.8){\includegraphics[scale=0.4, angle=90]{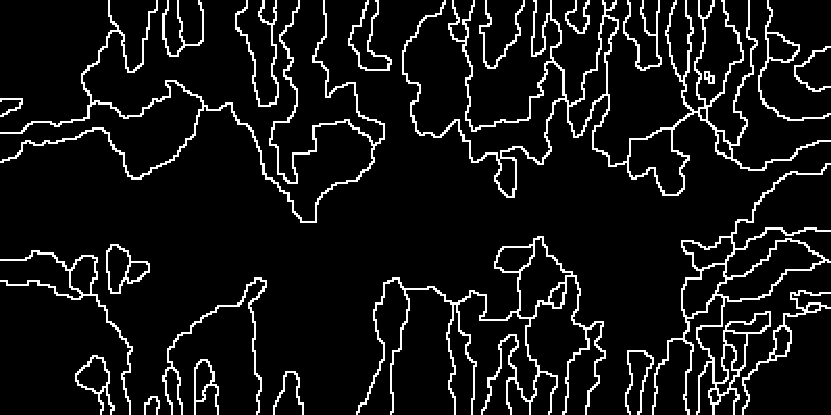}}
    \end{pspicture}}\qquad
  \subfloat[\scriptsize{Sobel method applied to (i).}]{\label{fig:12}
    \begin{pspicture}(0,0)(2.8,5.6)
      \rput(1.4,2.8){\includegraphics[scale=0.4, angle=90]{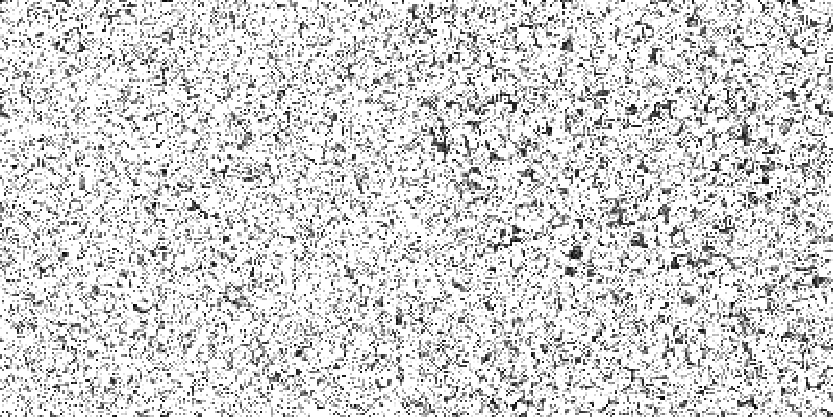}}
    \end{pspicture}}\\       
  \caption[]{Edge location comparison: the {\it gPb} method and the classical Sobel method coupled with the Lee filter.}
  \label{fig.comparison}
\end{center}
\end{figure*}

In Figure~\ref{fig.comparison} we compare the interval estimation approach proposed in this paper at the $95\%$ confidence level (panels~\ref{fig:2}, \ref{fig:6} and \ref{fig:10}) to the {\it gPb} estimator proposed in \cite{arbelaez} (panels~\ref{fig:3}, \ref{fig:7} and \ref{fig:11}) and also to Sobel's~\cite{sobel} approach after applying the speckle-reducing filter proposed by Lee~\cite{lee} (panels~\ref{fig:4}, \ref{fig:8} and \ref{fig:12}). Our goal is to verify whether the different methods are able to locate an edge that separates two regions of similar textures. We thus used simulated images. This was done so that we would have control over the true parameters. The edge is located between regions of equal sizes and textures are (1) extremely heterogeneous, $\alpha=-2$ and $\alpha=-3$ (panel~\ref{fig:1}), (2) heterogeneous, $\alpha=-7$ and $\alpha=-8$ (panel~\ref{fig:5}) and (3) homogeneous, $\alpha=-14$ and $\alpha=-15$ (panel~\ref{fig:9}). Here,  $L=1$ and $\gamma=-\alpha-1$. Each image has $200\times400$ pixels. In order to perform interval estimation, we divided each image into $10$ detection windows, each containing $20\times400$ pixels. In all situations, the point KW detector (red line) was able to locate the edge with high precision, if not perfectly 
(see panels~\ref{fig:2}, \ref{fig:6} and \ref{fig:10}). The interval lengths indicate that such detection is reliable when the regions the regions are extremely heterogeneous or homogenous. The confidence intervals obtained for heterogeneous regions are, however, wider and thus indicate that point edge estimation based on the KW detector may not be as reliable. The {\it gPb} detector yields a hierarchical representation for the edges in the image. Such representation is called Ultrametric Contour Map (UCM). The UCM at level $k$ yields a set of curves that are the segmentation edges at scale $k$. The scale is related to the probability  $P(x)$ that a given pixel $x$ lies on the edge. The white lines in panels~\ref{fig:3}, \ref{fig:7} and \ref{fig:11} are the pixels for which $P(x)\geq k$; here,  $k=0.07$. In all cases, no line emerged when we used $P(x)\geq 0.09$. Hence, each pixel of all images lies on an edge with very low probability according to the {\it gPb} detector. Additionally, none of the white lines fully agree with the true edge. Likewise, Sobel's method was not able to locate the edges; see panels~\ref{fig:4}, \ref{fig:8} and \ref{fig:12}. We obtained results that are similar to those obtained using Sobel's filter (not presented here) by using the gradient filter~\cite{gradiente} and also the method introduced by Touzi~\cite{touzi}. The {\it gPb} detector was computed using the {\tt Matlab} code available at \url{http://www.eecs.berkeley.edu/Research/Projects/CS/vision/grouping/resources.html}. The Lee filter was computed using the code available at {\tt Matlab Exchange} (\url{http://www.mathworks.com/matlabcentral/fileexchange}). Finally, other threshold based edges were computed using the {\tt Monteverdi-Orfeo Toolbox} (\url{http://www.orfeo-toolbox.org/otb/monteverdi.html}). 

\section{Concluding remarks}\label{S:conclusion}

We addressed the issue of edge detection in SAR images. As pointed out in Section~\ref{S:introduction}, the distributions of point edge estimators are usually unknown and can be estimated using data resampling. We used the bootstrap method to estimate the KW detector distribution. This allowed us to obtain confidence intervals for the edge location. Several bootstrap-based interval estimates were described and numerically evaluated. We addressed situations in which there is an edge and also situations in which there is none. The case of multiple edges will be addressed in future research.

Overall, the percentile confidence interval proved to be most reliable, especially in the challenging situation in which the regions on both sides of the edge have similar textures. The percentile interval typically displayed the best coverage but it is typically wider than alternative intervals. There is thus a trade-off between coverage and length. Percentile and basic bootstrap method delivered intervals with the smallest lengths. 

We proposed two variants of the bootstrap-$t$ method for edge interval estimation, denoted ST1 and ST2. They are less computationally costly than the standard Bootstrap-$t$ method. They tend to work equally well, ST1 being more computationally efficient. 

We have considered situations in which edge detection is carried out in an image region in which there is no edge. Point estimation will always locate an edge, even when there is none. In such situations, the resulting interval estimates tend to be fairly wide, thus signaling that the detected edge is not to be trusted. Very wide intervals can be taken as an indication that most likely there is no edge in that region of the SAR image and we can use this measure in an unsupervised system of edge detection. Real (not simulated) data were analyzed. We performed edge detection in several regions of a SAR image. 

It is important to remark that the edge interval estimators presented in this paper can be used with other types of image. Edge location in SAR images is particularly challenging due to the existence of speckled noise. Such noise can make the usual detection strategies to take noise for edges. Interval estimation can add useful information to the task of locating edges in such images. It can even signal that there is no edge in the region of the image under scrutiny. 

An interesting extension of the interval estimation methodologies considered in this paper involves edge detection in polarimetric SAR images, which are complex-valued. That would add important information to the edge point estimates that were proposed in the literature; see, e.g., \cite{schou} and \cite{frery2010}. We shall address that in future research.

\section*{Acknowledgements}

We gratefully acknowledge partial financial support from CAPES and CNPq. We thank Alejandro C.\ Frery for providing us with the data analyzed in Section~\ref{S:application}. Finally, we thank three anonymous referees for their thoughtful comments, suggestions and criticisms.

\end{document}